\input{aipcheck}

\documentclass[
  ]
  {aipproc}

\layoutstyle{8x11double}

\begin{document}

\title{Resonant Interactions Between Protons and Oblique Alfv\'en/Ion-Cyclotron Waves}


\author{Peera Pongkitiwanichakul}{
  address={Institute for the Study of Earth, Oceans and Space, University of New Hampshire, Durham NH,03824, USA}
}

\author{Benjamin D.G. Chandran}{
  address={Institute for the Study of Earth, Oceans and Space, University of New Hampshire, Durham NH,03824, USA}
}

\author{Philip A. Isenberg}{
  address={Institute for the Study of Earth, Oceans and Space, University of New Hampshire, Durham NH,03824, USA}
}

\author{Bernard J. Vasquez}{
  address={Institute for the Study of Earth, Oceans and Space, University of New Hampshire, Durham NH,03824, USA}
}

\begin{abstract}
  Resonant interactions between ions and Alfv\'en/ion-cyclotron (A/IC) waves may
  play an important role in the heating and acceleration of the fast solar
  wind. Although such interactions have been studied extensively for "parallel"
  waves, whose wave vectors~${\bf k}$ are aligned with the background magnetic
  field~${\bf B}_0$, much less is known about interactions
  between ions and oblique A/IC waves, for which the angle~$\theta$ between
  ${\bf k}$ and ${\bf B}_0$ is nonzero. In this paper, we present new numerical
  results on resonant cyclotron interactions between protons and oblique A/IC waves in
  collisionless low-beta plasmas such as the solar corona.
  We find that if some mechanism generates oblique high-frequency A/IC waves,
  then these waves initially modify the proton distribution function in such a
  way that it becomes unstable to parallel waves.  Parallel waves are then
  amplified to the point that they dominate the wave energy at the large
  parallel wave numbers at which the waves resonate with the particles. Pitch-angle scattering by these waves then causes the plasma to evolve towards
  a state in which the proton distribution is constant along a particular set of
  nested ``scattering surfaces'' in velocity space, whose shapes have been
  calculated previously. As the distribution function approaches this state, the
  imaginary part of the frequency of parallel A/IC waves drops continuously
  towards zero, but oblique waves continue to undergo cyclotron damping while simultaneously causing protons to
  diffuse across these kinetic shells to higher energies. We conclude that
  oblique A/IC waves can be more effective at heating protons than parallel A/IC
  waves, because for oblique waves the plasma does not relax towards a state in
  which proton damping of oblique A/IC waves ceases.
\end{abstract}

\maketitle

%%%%%%%%%%%%%%%%%%%%%%%%%%%%%%%%%%%%%%%%%%%%
%% MAINMATTER
%%%%%%%%%%%%%%%%%%%%%%%%%%%%%%%%%%%%%%%%%%%%

\section{Introduction}

Resonant interactions with Alfv\'en/ion-cyclotron (A/IC) waves are a
possible mechanism for ion heating in the solar corona, solar flares,
and the solar wind. Cyclotron heating in low-$\beta$ plasmas primarily
increases a particle's thermal motions perpendicular to the background
magnetic field~${\bf B}_0$~\cite{Hollweg02}, and may thus be
able to explain the observed temperature anisotropies of minor ions in
the solar corona~\cite{Kohl98} and protons in the fast
solar wind~\cite{Marsch2004}. (Here, $\beta = 8\pi p/B_0^2$, where $p$
is the plasma pressure.)  In solar flares, the magnetic tension in
reconnected magnetic field lines leads to large-scale flows that can
generate waves and turbulence. Wave energy is then transfered from
large scales to small scales by nonlinear wave-wave
interactions~\cite{Yan2004,cha05}. Small-scale A/IC waves may be
sufficiently energetic in solar flares to stochastically accelerate
ions to high energies~\cite{Miller95,Liu2004}.  Most
previous studies of ion heating by A/IC waves have focused on
``parallel waves,'' for which the angle~$\theta$ between the wave
vector~${\bf k}$ and~${\bf B}_0$ is zero. On the other hand, in the
solar corona and solar flares, A/IC wave intensities are not
restricted to~$\theta=0$. In this paper, we thus focus on resonant
interactions between protons and oblique A/IC waves, for which $\theta
\neq 0$.

\section{ Wave-Particle Interactions}

We consider A/IC waves in a low-$\beta$, proton-electron plasma, and assume that the real part of the wave frequency, $\omega_{kr}$, is given by the cold-plasma A/IC dispersion relation~\cite{Stix92},
\[
w^2 = \frac{k^2_n}{2}(1+\cos^2\theta)+\frac{k^4_n}{2}\cos^2\theta
\]
\begin{equation} 
- \frac{k^2_n}{2}\left [ k^4_n \cos^4\theta 
+2k^2_n\cos^2\theta(1+\cos^2\theta)+\sin^4\theta\right ]^{1/2}, 
\label{eq:disp0} 
\end{equation} 
where $w=\omega_{kr}/\Omega_p$, $\Omega_p$ is the proton cyclotron 
frequency, $k_n=kv_{\rm A}/\Omega_p$, 
and $v_{\rm A}=B_0/\sqrt{4 \pi\rho_0}$ is the Alfv\'en speed.
Protons strongly interact with such waves only when the resonance condition,
\begin{equation}
\omega_{kr}-k_{\|} v_\|=n\Omega_p,
\label{eq:res} 
\end{equation}
is satisfied, where $v_\parallel$ ($v_\perp$) is the component of the particle
velocity~${\bf v}$ parallel (perpendicular) to~${\bf B}_0$, $k_\parallel$
($k_\perp)$ is the component of~${\bf k}$ parallel (perpendicular) to~${\bf
  B}_0$, and $n$ is any integer~\cite{Kennel66a,Stix92}. The strongest interaction occurs for n=1~\cite{Hollweg02}.  

\begin{figure}
  \includegraphics[height=.24\textheight]{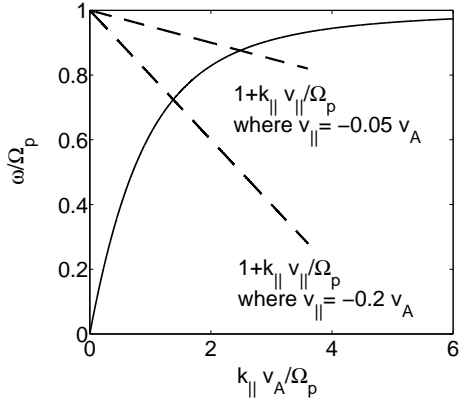}
  \caption{The intersections between lines satisfy the resonant condition n=1.}
  \label{resonant}
\end{figure}

Figure~\ref{resonant} plots $w(k_n)$ for $\theta=0$ (solid line), as well as two dashed lines
corresponding to $1 + k_\parallel v_\parallel/\Omega_p$ for two different values
of the proton parallel velocity~$v_\parallel$. The intersections of these two
lines and the $w(k_n)$ curve correspond to solutions of equation~(\ref{eq:res})
with~$n=1$.  When $|v_\parallel| \ll v_{\rm A}$, 
equations~(\ref{eq:disp0}) and (\ref{eq:res})  imply that~\cite{Hollweg99}
\begin{equation}
k_{n\parallel, {\rm res}} \simeq \left| \frac{v_\parallel}{v_{\rm A}}\right|^{-1/3},
\label{eq:h99} 
\end{equation} 
and 
\begin{equation}
v_{\rm ph}(v_\parallel) \simeq \pm v_{\rm A}^{2/3}|v_\parallel|^{1/3},
\label{eq:vph} 
\end{equation} 
where $k_{n\parallel, {\rm res}}$ is the value of $k_\parallel v_{\rm A}/\Omega_p$ that satisfies
equation~(\ref{eq:res}) for $\theta=0$, and $v_{\rm ph}(v_\parallel)$ is the parallel phase
velocity~$\omega/k_\parallel$ of the resonant waves at $\theta=0$.

\begin{figure}[h]
  \includegraphics[height=.35\textheight]{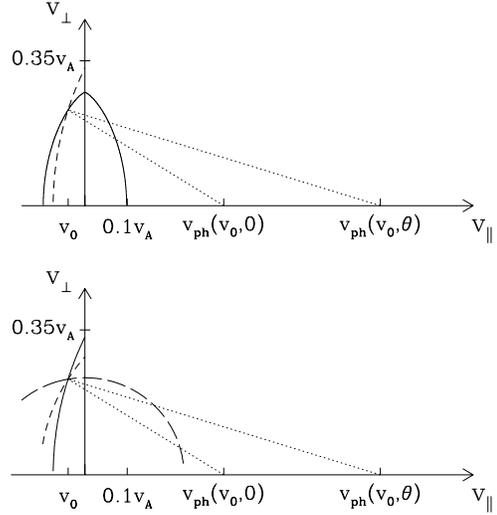}
\caption{\footnotesize {\em Upper panel:} The proton distribution is
  taken to be constant along the $\eta= \mbox{constant}$ scattering
  contours for waves with~$\theta = 0$, and to decrease as one moves
  to contours that are farther from the origin.  The solid line is the
  $\eta = 0.075 v_{\rm A}^2$ contour.  Protons interacting with
  oblique waves will diffuse upward along the dashed line, gaining
  energy, and damping the oblique waves. {\em Bottom panel:} The
  proton distribution function is now taken to be constant along the
  scattering contours for oblique waves with some nonzero~$\theta$,
  and to decrease as one moves to contours that are farther from the
  origin. The solid line illustrates one such contour.  Protons
  interacting with waves with~$\theta=0$ will diffuse down the density
  gradient along the short-dashed line, losing energy, and amplifying
  the $\theta=0$ waves.  The long-dashed line is a contour of constant
  energy in the plasma frame.
\label{fig:eta_surf2}}
\end{figure}

Resonant interactions between particles and waves cause particles to diffuse in
the $v_\parallel- v_\perp$ plane.  Particles interacting with a particular wave
with wave vector~${\bf k}$ and frequency~$\omega$ diffuse within the
$v_\parallel-v_\perp$ plane along a curve for which the particle energy is
conserved in a frame moving with velocity~$\omega/k_\parallel$ in the direction
of~${\bf B}_0$~\cite{Kennel66a,Stix92} (wave pitch-angle scattering). If protons interact with A/IC waves with a broad range
of~$k_\parallel$ values and~$\theta=0$, then particles will diffuse along closed
contours in the $v_\parallel-v_\perp$ plane. These contours are defined by the
equation $\eta = \mbox{ constant}$, where~\cite{Isenberg1996,Chandran2009}
\begin{equation}
\eta \simeq v_\perp^2 + (3/2) v_{\rm  A}^{2/3} |v_\parallel|^{4/3}.
\label{eq:eta3} 
\end{equation} 
If resonant wave-particle interactions control the evolution of the proton
distribution function~$f$, then $f$ becomes constant on surfaces of
constant~$\eta$.  Once $f=f(\eta)$, protons stop gaining or losing energy from
interacting with parallel A/IC waves, and parallel A/IC waves are neither damped
nor amplified~\cite{Kennel66a, Kennel66b}.  The shape of the
contour~$\eta = 0.075 v_{\rm A}^2$ is shown in the top panel of
Figure~\ref{fig:eta_surf2}.

When protons interact with oblique A/IC waves with a single nonzero
value of~$\theta$, they also undergo wave pitch-angle scattering along
a set of nested, closed contours in the $v_\parallel-v_\perp$
plane. However, at a fixed $v_\parallel \ll v_{\rm A}$, the parallel
phase velocity $\omega/k_\parallel$ of resonant oblique waves, denoted
$v_{\rm ph}(v_\parallel, \theta)$, is greater than the parallel phase
velocity of resonant parallel waves, $v_{\rm
  ph}(v_\parallel)$~\cite{Chandran2009}. Thus, at a fixed point in the
$v_\parallel - v_\perp$ plane, the oblique-wave scattering contour has
a larger slope than the parallel-wave scattering contour, as
illustrated in Figure~\ref{fig:eta_surf2}, which is adapted
from~\cite{Chandran2009}.  In the top panel of
Figure~\ref{fig:eta_surf2}, we take the proton distribution function
to be constant along the $\eta=\mbox{ constant}$ scattering contours
of $\theta=0$ waves, and the solid-line curve corresponds to $\eta =
0.075 v_{\rm A}^2$. Protons at $v_\parallel = v_0$ that are scattered
by oblique A/IC waves with some nonzero value of~$\theta$ will scatter
along the dashed-line trajectory, which corresponds to constant energy
as measured in a reference frame moving at velocity $v_{\rm
  ph}(v_0,\theta)$ along the magnetic field. [We have
artificially increased $v_{\rm ph}(v_0, \theta)$ relative to $v_{\rm
  ph}(v_0, 0)$ in both panels of Figure~\ref{fig:eta_surf2} to make
the figure easier to read.] If we take $f$ to be a decreasing function
of~$\eta$, then there will be a net diffusive flux of protons upward
along this dashed line, resulting in an increase in particle energy
and damping of oblique waves.  Elsewhere, we have calculated
analytically the damping rate of oblique A/IC waves assuming that
$f=f(\eta)$~\cite{Chandran2009}.

In the bottom panel of Figure~\ref{fig:eta_surf2}, we take the proton
distribution to be constant along the scattering contours corresponding to waves
with a single nonzero value of~$\theta$. One of these contours is now drawn with
a solid line. Protons at $v_\parallel = v_0$ interacting with A/IC waves with
$\theta = 0$ will scatter along the short-dashed line in this panel, which
locally corresponds to an $\eta = \mbox{constant}$ curve. If we take the proton
distribution to decrease as one moves to closed (solid-line) contours that are farther from
the origin, then there will be a net diffusive flux of protons downward along
the short-dashed line in the bottom panel of Figure~\ref{fig:eta_surf2}. In this case,
the protons will lose energy, and waves with $\theta =0$ will be amplified.

Based on these arguments, we make the following conjecture.  If some mechanism
generates high-frequency A/IC waves with a range of $\theta$ values, and if the
form and evolution of the proton distribution function are dominated by
wave-particle interactions, then interactions involving waves with
nonzero~$\theta$ will act to make the constant-$f$ contours steeper in the
$v_\parallel - v_\perp$ plane than the $\eta = \mbox{constant}$ scattering
contours of the parallel waves. This in turn will lead to the amplification of
waves with $ \theta = 0$ and cause the angular distribution of the waves at
large $k_{n}$ that resonate with the protons to become sharply peaked
around~$\theta = 0$.  Wave-particle interactions will then become dominated by
waves with~$\theta=0$, $f$ will become approximately constant along surfaces of
constant~$\eta$, and oblique waves will be damped. In the next section, we
describe numerical calculations that support this conjecture. 

\section{Numerical Calculations}

In the quasilinear theory of resonant wave-particle interactions, protons diffuse in velocity space as described by the equation
\begin{eqnarray}
\frac{\partial f}{\partial t}&=&\lim_{V \to \infty} \frac{\pi q^2}{4 m_p^2} \sum_{n=-\infty}^\infty \int d^3 \vec{k}\frac{(2\pi)^{-3}}{V} \frac{1}{v_\perp} G v_\perp \nonumber \\
&& \delta (\omega_{k r}-k_\| v_\|-n\Omega_p) | \psi_{n,k} | ^2 G f,
\label{eq:QLT0}
\end{eqnarray}
where $G=\left(1-k_\|v_\|/\omega_{k r}\right)\partial/\partial v_\perp
+(k_\|v_\perp/\omega_{k r})\partial/\partial v_\|$,
$\psi_{n,k}=E^+_kJ_{n+1}(k_\perp v_\perp/\Omega_p)+
E^-_kJ_{n-1}(k_\perp v_\perp/\Omega_p)$ (where we have set
$E_{k,z}=0$), $V$ is the volume, $J_n$ is the Bessel function of
order~$n$, $E^{\pm}_k=E_{kx} \pm i E_{ky} $, and ${\bf E}_k$ (${\bf
  B}_k$) is the Fourier transform of the electric (magnetic)
field~\cite{Kennel66a,Stix92}.  To integrate
equation~(\ref{eq:QLT0}) numerically, we discretize velocity space
using cylindrical coordinates with 100 grid cells spanning the
interval $0 < v_{\perp} < 0.5$ $v_{\rm A}$ and 100 grid cells for the
interval $-0.5$ $v_{\rm A} < v_\parallel < 0$. We discretize k-space
using spherical coordinates with 30 grid cells spanning the interval
$0 < k < 400$ $\Omega_p/v_{\rm A}$ and 10 grid cells for the interval
$0 < \theta< \pi/2$.  We assume cylindrical and reflectional symmetry.

The wave energy per unit volume in $k$ space is  $W_k=\left \{ \textbf{B}_k^* \cdot \textbf{B}_k +\textbf{E}_k^*\cdot [\partial (\omega \underline{\underline{\epsilon}}_{\,h})/\partial \omega]\cdot \textbf{E}_k \right\}/(8\pi) $, 
where $\underline{\underline{\epsilon}}_{\,h}$ is the hermitian part of the dielectric tensor~\cite{Stix92}. We evolve~$W_k$ in time using ``detailed energy conservation,'' i.e., by keeping track of the change in the particle energy~$\Delta {\cal E}$ resulting from waves in each wavenumber bin, and deducting $\Delta {\cal E}$ from the wave energy in that wavenumber bin.  It can be shown that this method is equivalent to evolving the waves using the analytic formula for the damping or growth rate~$\gamma_k$ given by~\cite{Kennel67} for the limit in which~$|\gamma_k| \ll |\omega_{kr}|$.

We  integrate equation~(\ref{eq:QLT0}) for~$f$ (and the corresponding equation for~$W_k$ resulting from detailed energy conservation) using an implicit time stepping algortihm, the biconjugate gradient stabilized method~\cite{Vorst03}.
 We hold the wave power spectrum~$W_k$ fixed at
the value $10^{-7}k^{-1} V B^2_0 v^2_{\rm A}/(8\pi\Omega_p^2)$ for
$\theta = (13/40) \pi$. At all other values of $\theta$, we initially
set $W_k=10^{-10}k^{-1} V B^2_0 v^2_{\rm A}/(8\pi\Omega_p^2)$ and then
we allow $W_k$ to vary in time. The proton distribution function $f$
is initially Maxwellian with a thermal speed $\sqrt{<v^2>}$ of $0.012$
$v_{\rm A}$. The minimum value of $\theta$ at the cell center in the
run that we present is~$\theta_{\rm min} = \pi/40$. We note that the
scattering contours for waves with $\theta = \theta_{\rm min}$ are
very similar to the $\eta = \mbox{ constant}$ lines.

\begin{figure}
  \includegraphics[height=.28\textheight]{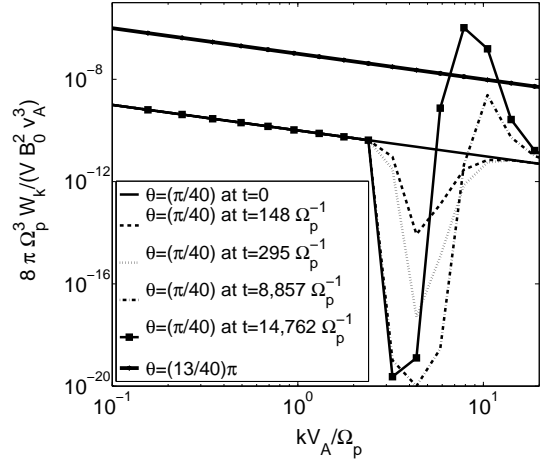}
  \caption{The power spectrum of the smallest-$\theta$ waves are shown at different times, and are compared to the fixed power spectrum of the waves with $\theta = 13\pi/40$.}
  \label{wave}
\end{figure}

\begin{figure}
  \includegraphics[height=.29\textheight]{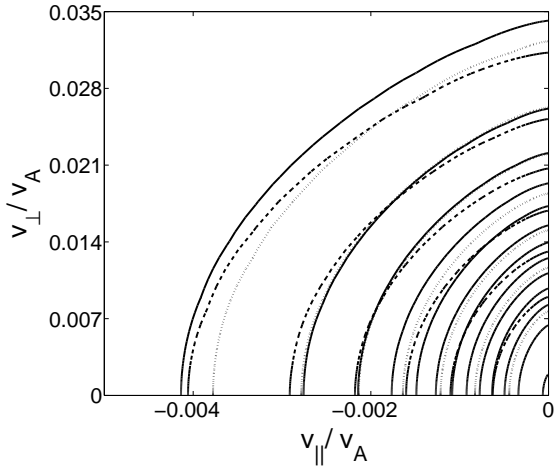}
  \caption{At $t=8,857 \Omega^{-1}_p$, the contours of constant~$f$ (solid lines) become steeper than the scattering
contours of the waves with $\theta = \pi/40$ (dashed lines) and almost aligned with the scattering contours of the waves with $\theta = (13/40)\pi$ (dotted lines). Waves with $\theta = \pi/40$ become unstable and subsequently grow in amplitude.}
  \label{c300}
\end{figure}

\begin{figure}
  \includegraphics[height=.29\textheight]{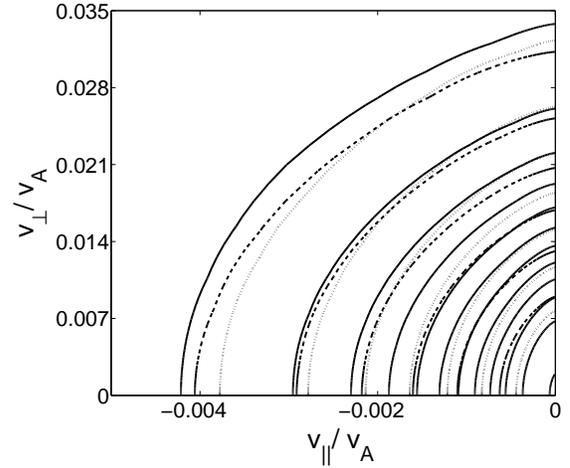}
  \caption{At $t=14,762 \Omega^{-1}_p$, the amplitudes of the waves with $\theta=\pi/40$ exceed the fixed amplitudes of the waves at $\theta = (13/40)\pi$ and the contours of constant~$f$ (solid lines) become almost aligned with the scattering contours of the waves with $\theta = \pi/40$ (dashed lines).}
  \label{c500}
\end{figure}

At early times, all the waves are damped by interacting with thermal particles
as shown in Figure \ref{wave}.  Waves with $\theta=(13/40)\pi$, which are
initially dominant, cause particles to diffuse along the relatively steep
scattering contours shown by the dotted lines in Figure~\ref{c300}.  When
the contours of constant~$f$ in the simulation become steeper than the
scattering contours of the waves with $\theta = \theta_{\rm min}$, waves with
$\theta=\theta_{\rm min}$ are amplified. After the energy in waves with $\theta=\theta_{\rm min}$
exceeds the energy in waves with $\theta = (13/40)\pi$, the small-$\theta$ waves
begin to dominate, and the contours of constant~$f$ start to align with the
scattering contours of the waves with~$\theta = \theta_{\rm min}$, as shown in
Figure~\ref{c500}. Even though we ongoingly input energy only into oblique waves, (quasi) parallel waves at $\theta = \theta_{\rm min}$
ultimately dominate, and the distribution function evolves to a state in which $f \simeq f(\eta)$ 
%in the region in velocity space that contains most of the particles.

\section{Discussion}

Isenberg~\cite{Isenberg2004} has shown that  A/IC waves with $\theta = 0$ are unable to explain the heating and acceleration of protons in the fast solar wind, primarily because~$f$ relaxes towards a state in which $f\simeq f(\eta)$, after which the protons are only weakly heated by the
waves. In contrast, we have shown that when protons are heated by oblique A/IC waves, the distribution function does not relax towards a state in which oblique-wave damping vanishes. Instead, $f$ again approaches a state in which~$f=f(\eta)$, and oblique waves continue to damp on the protons, causing protons to diffuse across $\eta=\mbox{ constant}$ surfaces in the $v_\perp-v_\parallel$ plane. Because of this, oblique A/IC waves have the potential to be more effective than $\theta=0$ waves at heating protons in the corona and solar wind.

\begin{theacknowledgments}

\end{theacknowledgments}

This work was supported by NSF Grants ATM-0851005 and 0850205, by DOE
under Grant DE-FG02-07-ER46372, by NSF-DOE Grant AST-0613622, and by
NASA under Grants NNX07AP65G and NNX08AH52G.

\bibliographystyle{aipproc}   % if natbib is available

\bibliography{}

\IfFileExists{\jobname.bbl}{}
 {\typeout{}
  \typeout{******************************************}
  \typeout{** Please run "bibtex \jobname" to optain}
  \typeout{** the bibliography and then re-run LaTeX}
  \typeout{** twice to fix the references!}
  \typeout{******************************************}
  \typeout{}
 }

%%%%%%%%%%%%%%%%%%%%%%%%%%%%%%%%%%%%%%%%%%%
%% The following lines show an example how to produce a bibliography
%% without the help of the BibTeX program. This could be used instead
%% of the above.
%%%%%%%%%%%%%%%%%%%%%%%%%%%%%%%%%%%%%%%%%%%

\end{document}